\documentclass[aps,preprint,showpacs]{revtex4}
\begin{document}        %
\draft
\title{Comments on "Effects of wall roughness on flow in nanochannels"}
\author{Zotin K.-H. Chu} 
\affiliation{3/F, 24, 260th. Lane, First Section, Muja Road,
Taipei, Taiwan 116, China}
\begin{abstract}
We make remarks on  Sofos {\it et al.}'s [{\it Phys. Rev. E} 79,
026305 (2009)] paper. The focus is about the monotonicity of the
slip length of which it is different from previous similar
numerical simulation. We also offer a possible explanation for
this.
%
\end{abstract}
\maketitle
\bibliographystyle{plain}
Sofos {\it et al.} just showed that [1] the maximum value of
streaming velocity in the center of the nanochannel is not
significantly affected by the presence of roughness. Meanwhile as
the rectangular wall cavities become narrower (as the $p$ value
increases) velocity values inside the cavities decrease and fluid
atoms tend to be trapped inside them. With above results, Sofos
{\it et al.} observed that slip on the boundary diminishes as
fluid atoms are trapped inside the cavities [1]. Sofos also noted
that they don't have a monotonic behavior for the maximum velocity
values as $p$ increases from $p=0$ to $6$, but they concluded that
all maximum velocity values are smaller in the rough channel cases
compared to a smooth one. Note that in [1] an external driving
force $F_{ext}=0.013 44 \,\epsilon/\sigma$ ($\sigma=0.3405$ nm,
$\epsilon=119.8 ^{\circ}$K) is applied along the $x$ direction to
drive the flow with the temperature being kept to be constant at
$T^*=1$ ($\epsilon/k_B$, $k_B$ is Boltzmann's constant) and with
the application of Nos\'{e}-Hoover thermostats.\newline Firstly,
for $p=0$ (smooth) case, there is a strange (largest) peak for the
total average number density ($N^*$) profiles as evidenced in Fig.
3 of [1]. There is no mathematical definition for $N^*$ in [1]?
The authors of [1] didn't explain this behavior ($N^* \sim 5$) or
pay specific attention to the Fig. 3? Is this strange peak due to
the smearing discontinuity or singularity occurred at the initial
step (during the numerical evolution) or near the outer boundary
(supposed to be a vacuum/matter interface [2] which is a sudden
jump)?
\newline Meanwhile, the trend of results, say, Fig. 8 in [1], is
different from that of previous results for similar geometry, say,
Fig. 4 (a) in [3] or Fig. 7 in [4] (the role of $p$ in [1] is
similar to that of $ka$ in [4]). To be precise, the slip behavior
in the latter is monotonic while that in the former is not
monotonic. Note that the (numerical) simulation step for the
system is $t=0.005 \tau$ ($\tau$ is in units of $\sqrt{m \sigma^2
/\epsilon}$) which is the same as that in [4].
\newline To examine what happens for $p=2,3,$ and $6$ in [1] is
crucial to our understanding of the difference between the former
and the latter. The possible reasoning might be due to the authors
of [1] adopting this approach : Wall atoms are bound on fcc sites
and remain in their original positions (via an elastic spring
force ${\bf F}$ [1]). Meanwhile, the cavity for $p=3$ is of square
shape while that of $p=2$ (and, $p=6$, too) is of rectangular
shape. The roughness amplitude is about 10\% of the channel width
($\approx 2 \sigma$).  The combination of specific wall spring
forcing and square cavity thus makes the slip length at the rough
wall for $p=3$ case is a little bit larger than that of
$p=2$.\newline Finally, the present author likes to argue that as
there is a friction at the atomic scale or dissipation for the
flow driven by an external forcing (cf. $F_{ext}$ in [1]) along
the $x$-direction then the thermal problem for the presentation in
[1] cannot be neglected. It means once the dissipation occurs what
happens to the heating of fluid atoms as well as wall atoms? Can
we still fix the wall temperature (kept to be a constant $T^*=1$
in [1])? It is possible that the strange peak of $N^*$ mentioned
above might be due to the unbalanced heating (as the wall
temperature should be fixed) along the outer boundary since the
excess heat cannot be transferred outside.

\end{document}